\documentclass{llncs}
\usepackage{graphicx}
\usepackage{subfig}
\usepackage{multirow}
\usepackage{color,url}
\begin{document}
\title{Genome-wide Discovery of Modulators of Transcriptional Interactions
in Human B Lymphocytes}
\author{Kai Wang\inst{1,2} \and Ilya Nemenman\inst{2} \and
Nilanjana Banerjee \inst{2}\and Adam A. Margolin \inst{1,2} \and
Andrea Califano\inst{1,2,3}\textsuperscript{,}\thanks{Correspondence
should be addressed to \email{califano@c2b2.columbia.edu}}}
\institute{Department of Biomedical Informatics, Columbia
University\\
622 West 168th Street, Vanderbilt Clinic 5th Floor, New York, New York 10032\\
\and Joint Centers for Systems Biology, Columbia University\\
1130 St. Nicholas Ave, Rm 801A, New York, New York 10032\\
\and Institute of Cancer Genetics, Columbia University\\ Russ Berrie
Pavilion, 1150 St. Nicholas Ave, New York, 10032}

\maketitle

\begin{abstract}
Transcriptional interactions in a cell are modulated by a variety of
mechanisms that prevent their representation as pure pairwise
interactions between a transcription factor and its target(s). These
include, among others, transcription factor activation by
phosphorylation and acetylation, formation of active complexes with
one or more co-factors, and mRNA/protein degradation and
stabilization processes.\\
This paper presents a first step towards the systematic, genome-wide
computational inference of genes that modulate the interactions of
specific transcription factors at the post-transcriptional level.
The method uses a statistical test based on changes in the mutual
information between a transcription factor and each of its candidate
targets, conditional on the expression of a third gene. The approach
was first validated on a synthetic network model, and then tested in
the context of a mammalian cellular system. By analyzing 254
microarray expression profiles of normal and tumor related human B
lymphocytes, we investigated the post transcriptional modulators of
the MYC proto-oncogene, an important transcription factor involved
in tumorigenesis. Our method discovered a set of 100 putative
modulator genes, responsible for modulating 205 regulatory
relationships between MYC and its targets. The set is significantly
enriched in molecules with function consistent with their activities
as modulators of cellular interactions, recapitulates established
MYC regulation pathways, and provides a notable repertoire of novel
regulators of MYC function. The approach has broad applicability and
can be used to discover modulators of any other transcription
factor, provided that adequate expression profile data are
available.
\end{abstract}

\section{INTRODUCTION}

The reverse engineering of cellular networks in prokaryotes and lower
eukaryotes~\cite{kw:Fri2004,kw:Gar2003}, as well as in more complex
organisms, including mammals~\cite{kw:Elk2003,kw:Stu2003,kw:Bas2005},
is unraveling the remarkable complexity of cellular interaction
networks. In particular, the analysis of targets of specific
transcription factors (TF) reveals that target regulation can change
substantially as a function of key modulator genes, including
transcription co-factors and molecules capable of post-transcriptional
modifications, such as phosphorylation, acetylation, and
degradation. The yeast transcription factor STE12 is an obvious
example, as it binds to distinct target genes depending on the
co-binding of a second transcription factor, TEC1, as well as on the
differential regulation by MAP kinases FUS3 and
KSS1~\cite{kw:Zei2003}. Although the conditional, dynamic nature of
cellular interactions was recently studied in
yeast~\cite{kw:Lus2004,kw:Seg2003,kw:deL2005,kw:Pe'2002}, methods to
identify a genome-wide repertoire of the modulators of a specific
transcription factor are still lacking.

\begin{figure}[t]
    \centering
    \subfloat[ ]{
        \label{fig:fig1:a}
        \includegraphics[width=0.25\linewidth]{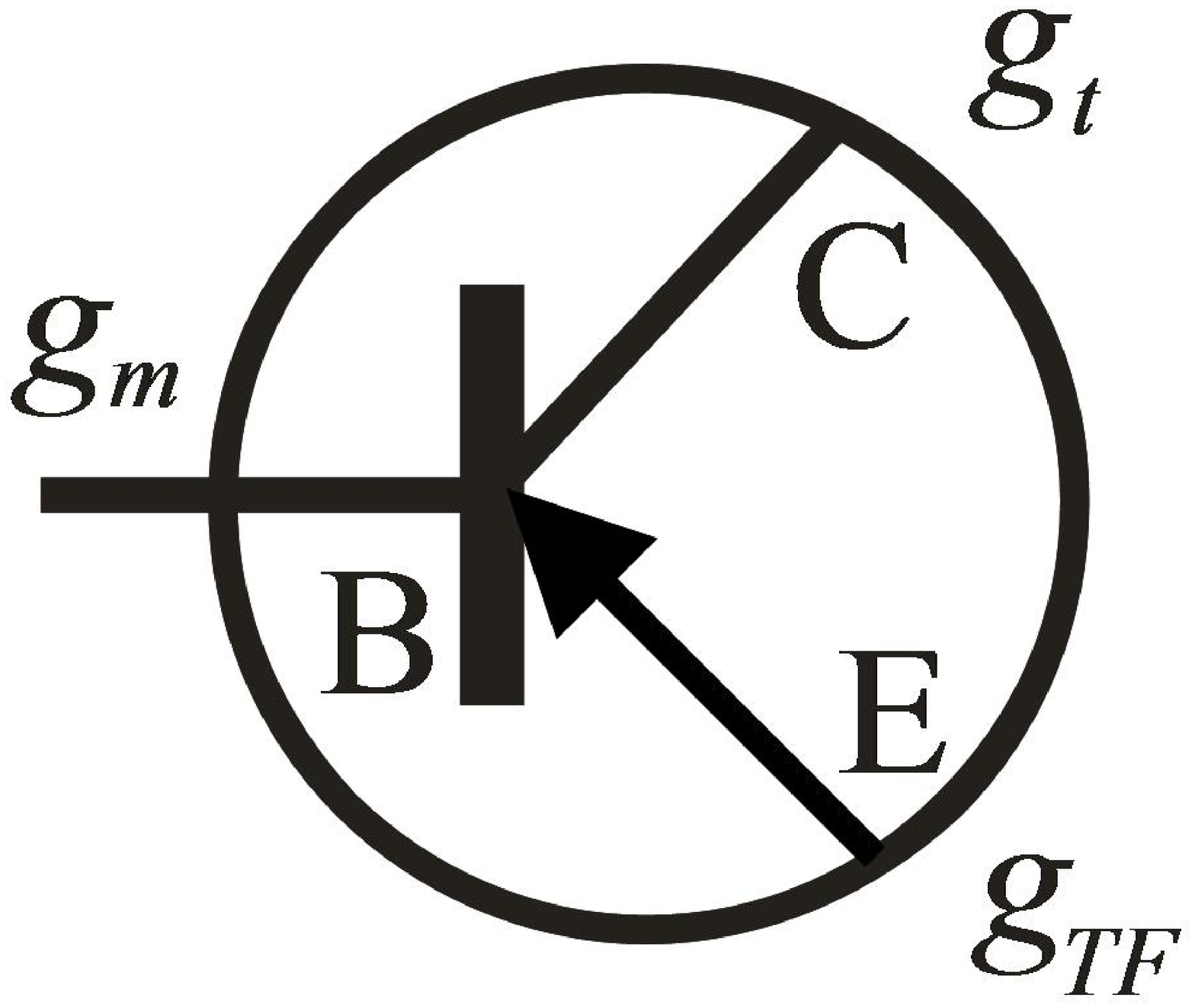}}
        \hspace{0.1\linewidth}
    \subfloat[ ]{
        \label{fig:fig1:b}
        \includegraphics[width=0.5\linewidth]{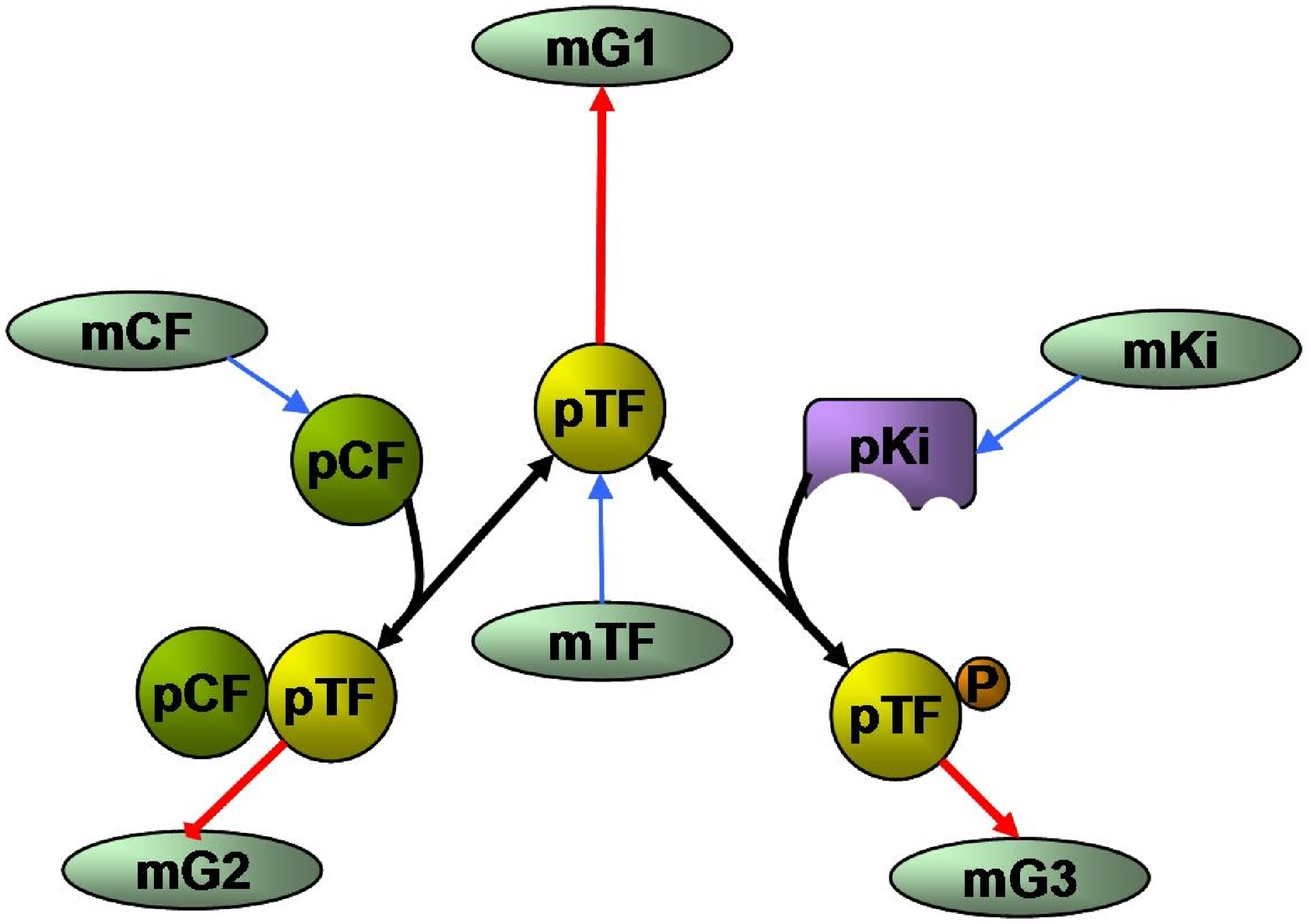}} \\[20pt]
    \subfloat[ ]{
        \label{fig:fig1:c}
        \begin{tabular}{|c|c|c|c|c|c|c|c|}
        \hline
        TF Interaction & $I$ & $I_{Ki}^-$ & $I_{Ki}^+$ & $\Delta I_{Ki}$ &
        $I_{CF}^-$ & $I_{CF}^+$ & $\Delta I_{CF}$ \\
        \hline
        Ki  &  --   &  --   &  --   &  --   & 0.007 & 0.016 & 0.009 \\
        \hline
        CF  &  --   & 0.001 & 0.018 & 0.008 &  --   & --    &  --   \\
        \hline
        $G_1$  & \textcolor{red}{0.732} & \textcolor{red}{0.542} & \textcolor{red}{0.579} &
        0.037 & \textcolor{red}{0.552} & \textcolor{red}{0.548} & 0.004 \\
        \hline
        $G_2$  & 0.064 & 0.079 & 0.093 & 0.014 & 0.007 & \textcolor{red}{0.378} &
        \textcolor{red}{0.371} \\
        \hline
        $G_3$  & 0.097 & 0.007 & \textcolor{red}{0.351} & \textcolor{red}{0.344} & 0.071 & 0.042 & 0.029 \\
        \hline
    \end{tabular}}
    \caption{Synthetic network model of transistor-like regulatory logic.
    (a) Transistor model.
    (b) Schematic representation of the synthetic network.
    (c) Unconditional MI, conditional MI and conditional MI difference for
    the TF interactions conditioning on the expression level of the
    Ki and the CF; entries colored in red are determined to be statistically
    significant.}
    \label{fig:fig1}
\end{figure}

In this paper, we explore a particular type of ``transistor like''
logic, shown in Fig.~\ref{fig:fig1:a}, where the ability of a
transcription factor $g_{\textit{\tiny{TF}}}$ (emitter) to regulate a
target gene $g_t$ (collector) is modulated by a third gene $g_m$
(base), which we shall call a modulator. Pairwise analysis of mRNA
expression profiles will generally fail to reveal this complex picture
because $g_m$ and $g_{\textit{\tiny{TF}}}$ (e.g., a kinase and a
transcription factor it activates) are generally statistically
independent and because the correlation between the expression of
$g_{\textit{\tiny{TF}}}$ and $g_t$ is averaged over an entire range of
values of $g_m$ and thus significantly reduced. However, we show that
by conditioning on the expression of the modulator gene (e.g., an
activating kinase), a statistically significant change in the
$g_{\textit{\tiny{TF}}} \leftrightarrow g_{t}$ correlation can be
measured, thus directly identifying key post-transcriptional
regulation mechanisms, including modifications by signaling molecules,
co-factor binding, chromatin accessibility, modulation of protein
degradation, etc. An important element of this analysis is that, while
signaling proteins are conventionally viewed as constitutively
expressed, rather than transcriptionally modulated, in practice their
abundance within a cell population is subject to fluctuations (either
functional or stochastic). Depending on the number of available
microarray expression profiles and on the range of fluctuation, this
may be sufficient to establish a $g_{\textit{\tiny{TF}}}
\leftrightarrow g_t$ statistical dependency, conditional on the
availability of one or more signaling
molecules.

We validated the approach on a simple synthetic network and then
applied it to the identification of key modulators of MYC, an
important TF involved in tumorigenesis of a variety of lymphomas. We
identify a set of 100 putative modulators, which is significantly
enriched in genes that play an obvious post-transcriptional or
post-translational modulation role, including kinases,
acyltransferases, transcription factors, ubiquitination and mRNA
editing enzymes, etc. Overall, this paper introduces the first
genome-wide computational approach to identify genes that modulate the
interaction between a TF and its targets. We find that the method
recapitulates a variety of known mechanisms of modulation of the
selected TF and identifies new interesting targets for further
biochemical validation.

\section{METHOD}

As discussed in~\cite{kw:Mar2005,kw:Nem2004}, the probability
distribution of the expression state of an interaction network can be
written as a product of functions of the individual genes, their
pairs, and higher order combinations. Most reverse engineering
techniques are either based on pairwise
statistics~\cite{kw:Bas2005,kw:Mar2005,kw:But2000}, thus failing to
reveal third and higher order interactions, or attempt to address the
full dependency model~\cite{kw:Fri2000}, making the problem
computationally untractable and under-sampled. Given these
limitations, in this paper we address a much more modest task of
identifying the ``transistor-like'' modulation of specific regulatory
interaction, a specific type of third order interactions that is
biologically important and computationally tractable in a mammalian
context. Furthermore, given the relatively high availability of
microarray expression profile data, we restrict our analysis to only
genes that modulate transcriptional interactions, i.e., a TF
regulating the expression of its target gene(s).

In our model, just like in an analog transistor where the voltage on
the base modulates the current between the other terminals, the
expression state of the modulator, $g_m$, controls the statistical
dependence between $g_{\textit{\tiny{TF}}}$ and $g_t$, which may range
from statistically independent to strongly correlated. If one chooses
mutual information (MI) to measure the interaction strength
(see~\cite{kw:Mar2005} for the rationale), then the monotonic
dependence of $I( g_{\textit{\tiny{TF}}}, g_t | g_m )$ on $g_m$, or
lack thereof, can reveal respectively the presence or the absence
of such a transistor-like interaction.

Analysis along the lines of~\cite{kw:Mar2005} indicates that currently
available expression profile sets are too small to reliably estimate
$I( g_{\textit{\tiny{TF}}}, g_t | g_m )$ as a function of $g_m$. To
reduce the data requirements, one can discretize $g_m$ into well
sampled ranges $g_m^i$. Then, $| I( g_{\textit{\tiny{TF}}}, g_t |
g_m^{i_1} ) - I( g_{\textit{\tiny{TF}}}, g_t | g_m^{i_2} ) | > 0$ (at
the desired statistical significance level) for any range pair $(i_1,
i_2)$ is a sufficient condition for the existence of the transistor
logic, either direct (i.e., $g_m$ is causally associated with the
modulation of the TF targets) or indirect (i.e., $g_m$ is co expressed
with a true modulator gene). Below we present details of an algorithm
that, given a TF, explores all other gene pairs $(g_m, g_t)$ in the
expression profile to identify the presence of the transistor logic
between the three genes.

\subsection{Selection of candidate modulator genes}

Given a expression profile dataset with $N$ genes and an a-priori
selected TF gene $g_{\textit{\tiny{TF}}}$, an initial pool of
candidate modulators $g_m$, $\{m\} \in 1,2,\dots,M$, is selected from
the $N$ genes according to two criteria: (a) each $g_m$ must have
sufficient expression range to determine statistical dependencies, (b)
genes that are not statistically independent of
$g_{\textit{\tiny{TF}}}$ (based on MI analysis) are excluded. The
latter avoids reducing the dynamic range of $g_{\textit{\tiny{TF}}}$
due to conditioning on $g_m$, which would unnecessarily complicate the
analysis of significance of the conditional MI change. It also removes
genes that transcriptionally interact with $g_{\textit{\tiny{TF}}}$,
which can be easily detected by pair-wise co-expression analysis, and
thus are not the focus of this work. We don't expect this condition to
substantially increase the false negative rate. In fact, it is
reasonable to expect that the expression of a post-transcriptional
modulator of a TF function should be statistically independent of the
TF's expression. For instance, this holds true for many known
modulators of MYC function (including MAX, JNK, GSK, and NF$\kappa$B).

Each candidate modulator $g_m$ is then used to partition the
expression profiles into two equal-sized, non-overlapping subsets,
$L_m^+$ and $L_m^-$, in which $g_m$ is respectively expressed at its
highest ($g_m^+$) and lowest ($g_m^-$) levels. The conditional MI,
$I^{\pm}=I( g_{\textit{\tiny{TF}}}, g_t | g_m^{\pm} )$, is then
measured as $I( g_{\textit{\tiny{TF}}}, g_t )$ on the subset
$L^{\pm}$. Note that this partition is not intended to identify the
over  or under expression of the modulator, but rather to estimate
$g_m^i$. Then, $| I( g_{\textit{\tiny{TF}}}, g_t | g_m^+ ) - I(
g_{\textit{\tiny{TF}}}, g_t | g_m^- ) | > 0$ for target genes using
the two tails of the modulator's expression range. The size of
$L_m^{\pm}$ is constrained by the minimal number of samples required
to accurately measure MI, as is discussed in~\cite{kw:Mar2005}.
Mutual information is estimated using an efficient Gaussian kernel
method on rank-transformed data, and the accuracy of the measurement
is known~\cite{kw:Mar2005}.

\subsection{Conditional mutual information statistics}

\begin{figure}[htb]
    \centering
    \subfloat[ ]{
        \label{fig:fig2:a}
        \includegraphics[width=0.48\linewidth]{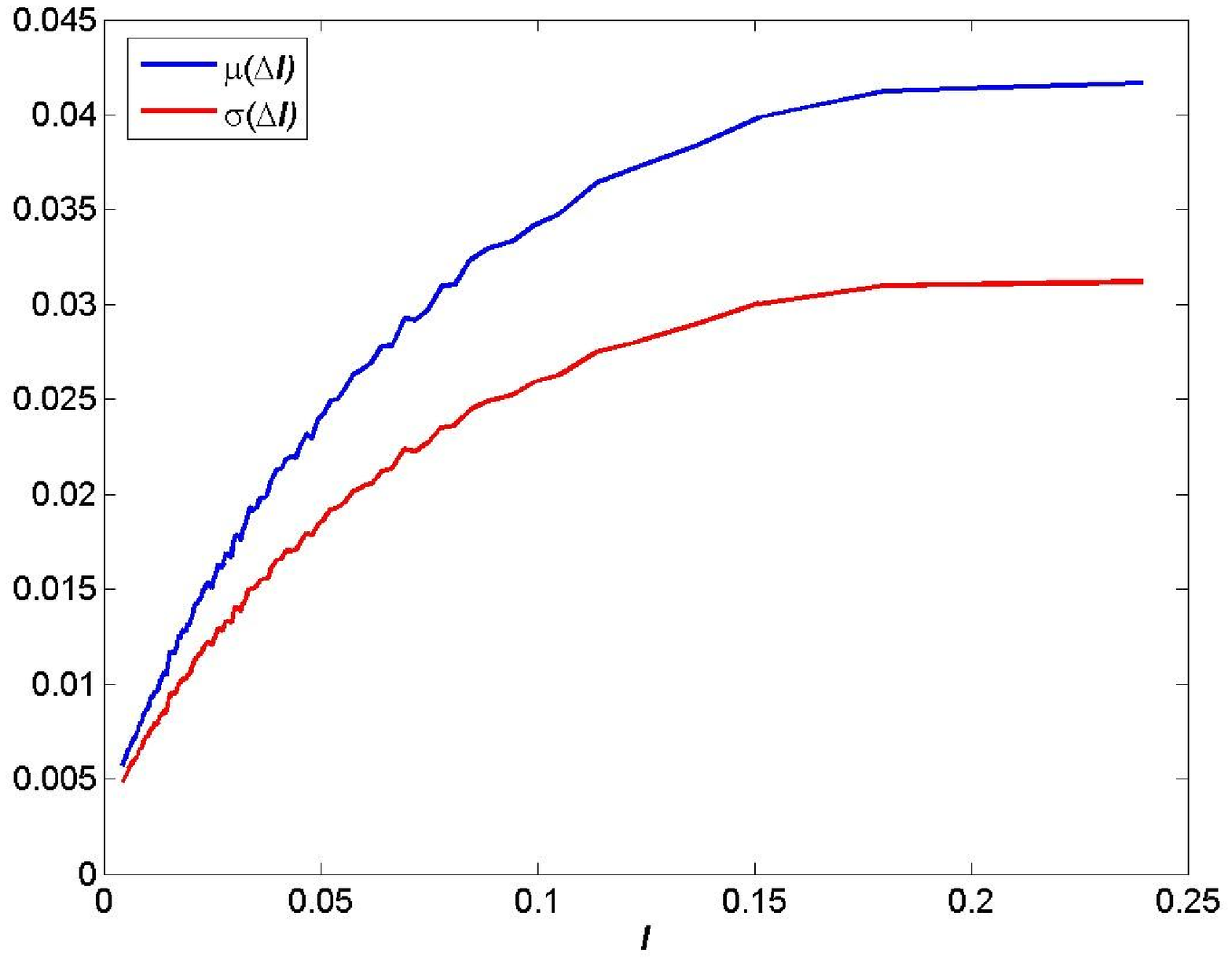}}
    \subfloat[ ]{
        \label{fig:fig2:b}
        \includegraphics[width=0.48\linewidth]{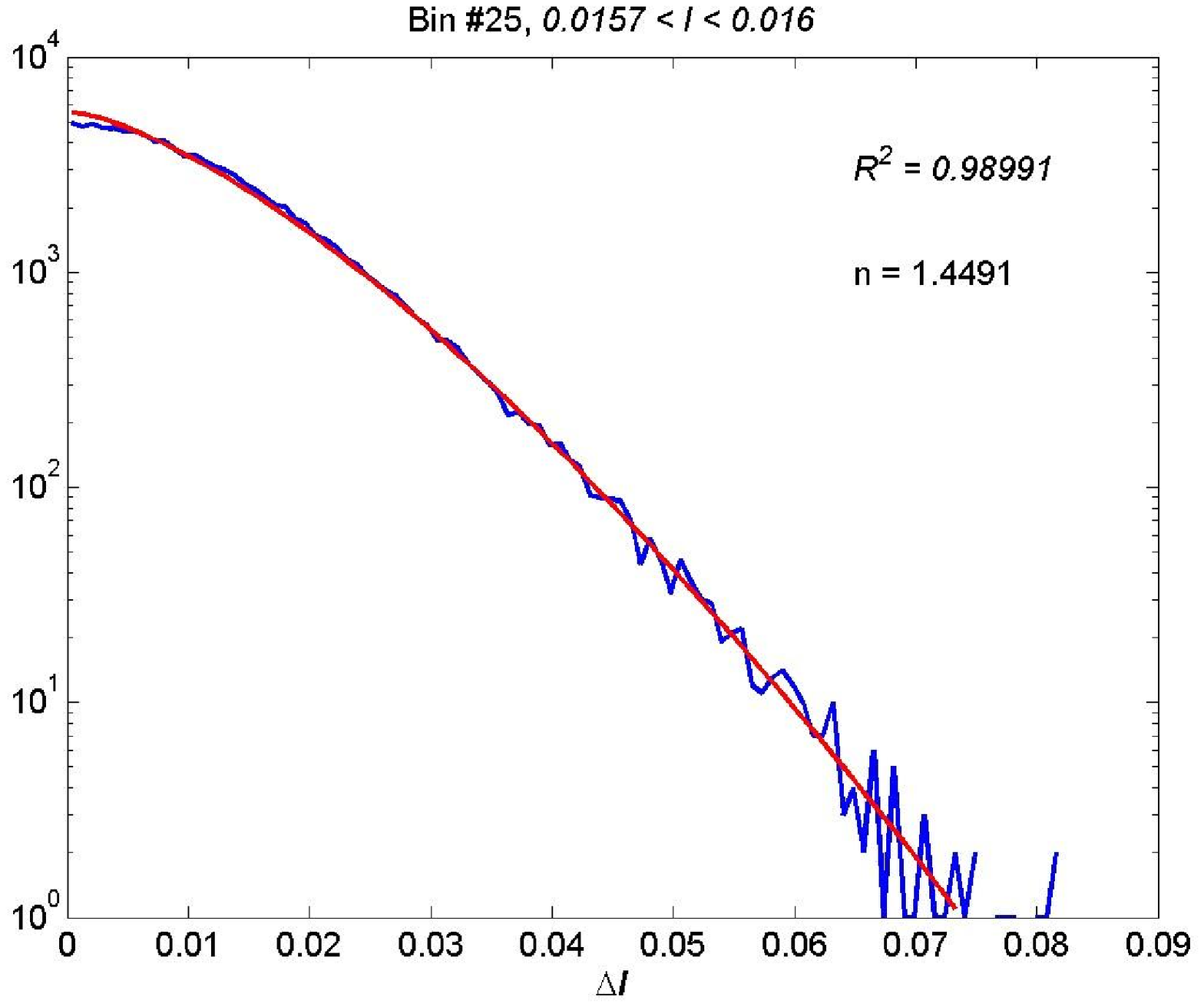}} \\
    \subfloat[ ]{
        \label{fig:fig2:c}
        \includegraphics[width=0.48\linewidth]{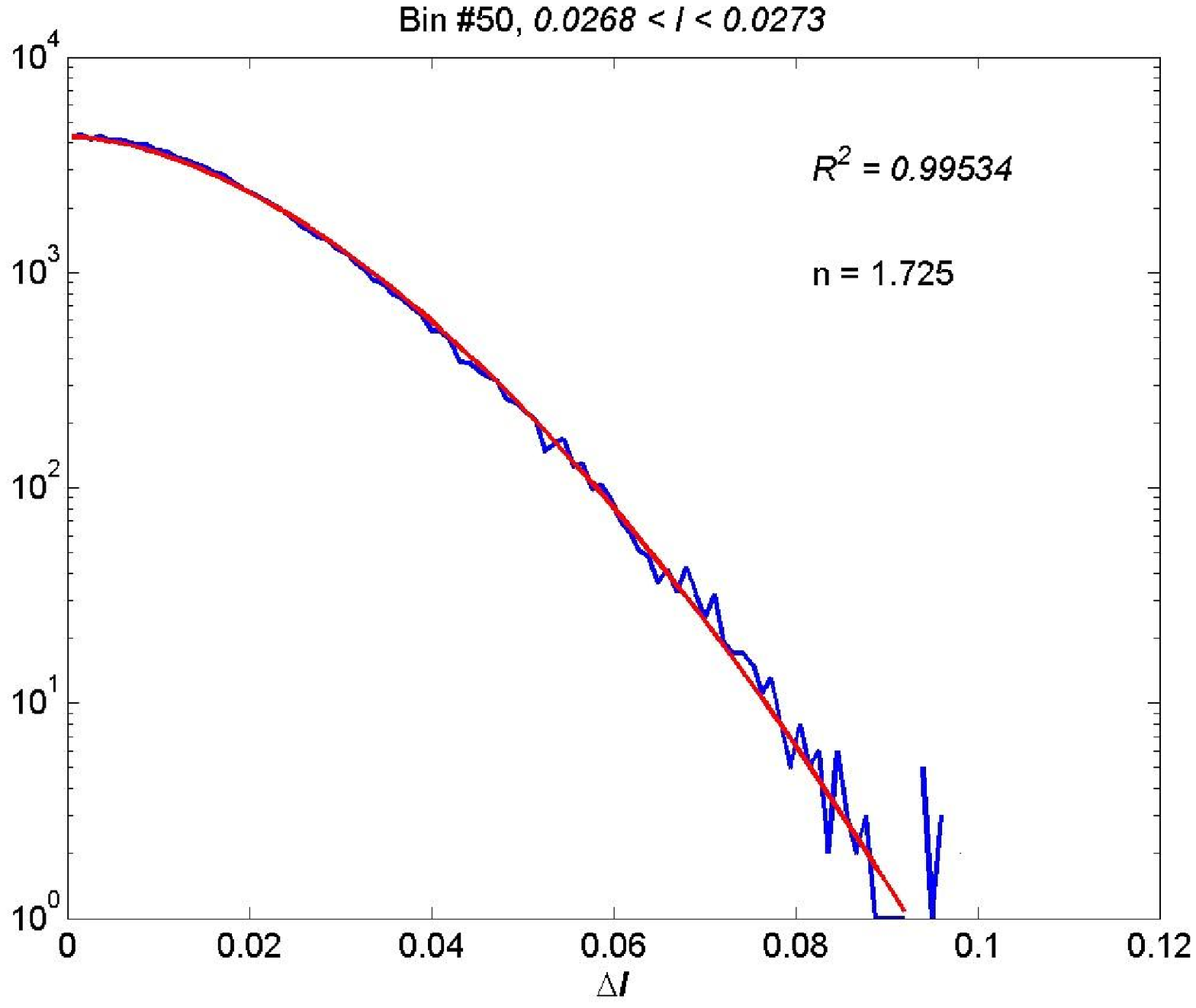}}
    \subfloat[ ]{
        \label{fig:fig2:d}
        \includegraphics[width=0.48\linewidth]{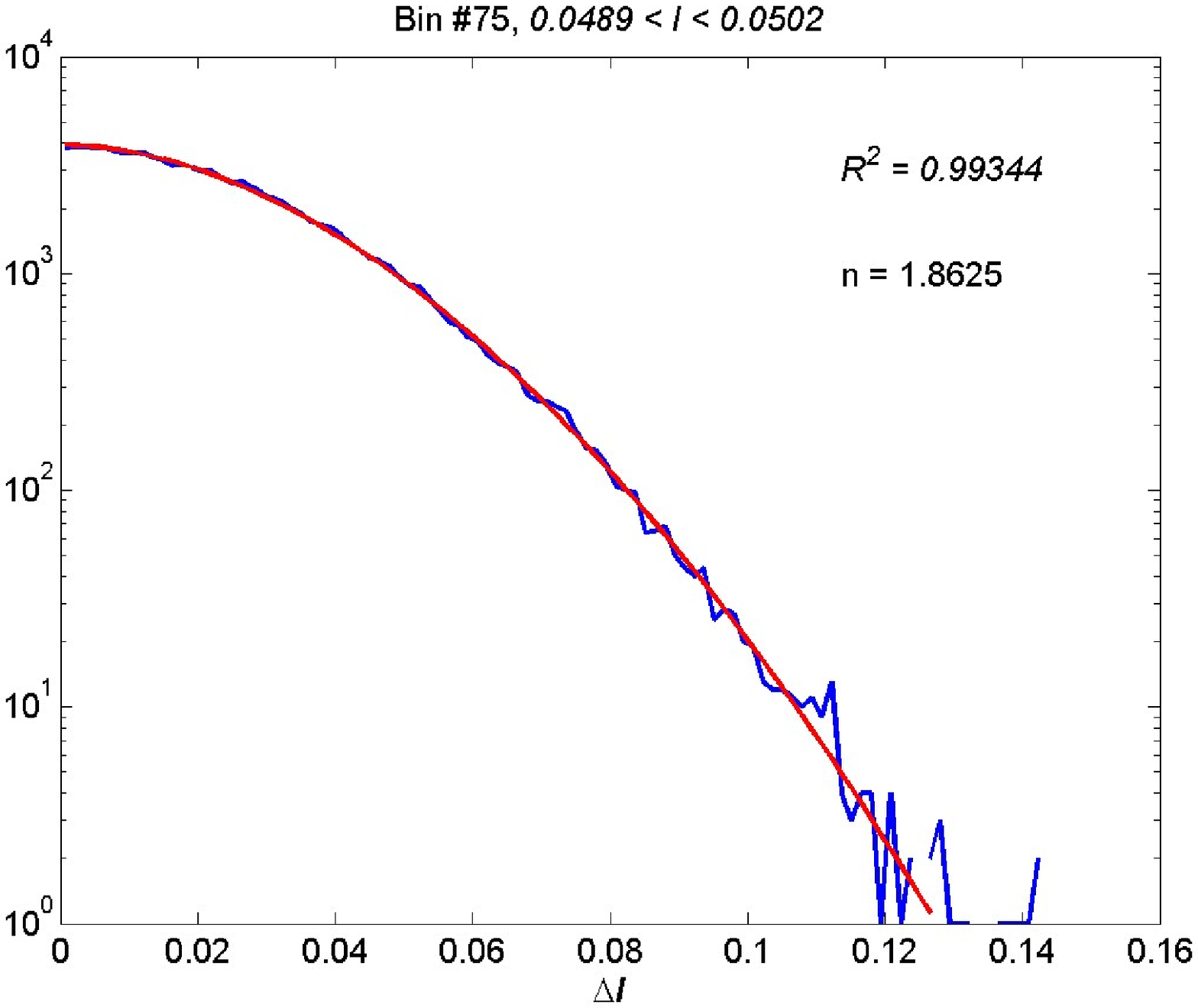}}
    \caption{Null distribution for the $\Delta I$ statistics.
    (a) Mean ($\mu$) and standard deviation ($\sigma$) of the $\Delta I$
    statistics in each bin as a function of $I$.
    (b) - (d), distribution of the $\Delta I$ statistics (blue curves)
    and the extended exponential function,
    $p(\Delta I)=exp(-\alpha \Delta I^n + \beta)$
    (red curves), obtained by least square fitting in bin 25, 50 and 75;
    a goodness-of-fit measure, $R^2$, and the value of $n$ are also
    shown for each bin.}
    \label{fig:fig2}
\end{figure}

Given a triplet $(g_m,g_{\textit{\tiny{TF}}},g_t)$, we define the
conditional MI difference as:
\begin{equation}
\Delta I( g_{\textit{\tiny{TF}}}, g_t | g_m ) = | I^+ - I^- | = |I(
g_{\textit{\tiny{TF}}}, g_t | g_m^+) - I( g_{\textit{\tiny{TF}}},
g_t | g_m^-)|
\end{equation}
For simplicity, hereafter we use $I$ for the unconditional MI (i.e.,
the MI across all samples) and $\Delta I$ for conditional MI
difference. To assess the statistical significance of a $\Delta I$
value, we generate a null hypothesis by measuring its distribution
across $10^4$ distinct ($g_{\textit{\tiny{TF}}}, g_t$) pairs with
random conditions. That is, for each gene pair, the non-overlapping
subsets $L_m^{\pm}$ used to measure $I^{\pm}$ and $\Delta I$ are
generated at random rather than based on the expression of a candidate
modulator gene (1000 $\Delta I$ from random sub-samples are generated
for each gene pair). Since the statistics of $\Delta I$ should depend
on $I$, we binned $I$ into 100 equiprobable bins, resulting in 100
gene pairs and $10^5 \Delta I$ measurements per bin. Within each bin,
we model the distribution of $\Delta I$ as an extended exponential,
$p(\Delta I)=exp(-\alpha \Delta I^n + \beta)$, which allows us to
extrapolate the probability of a given $\Delta I$ under this null
hypothesis model. As shown in Fig.\ref{fig:fig2}, both the mean and
the standard deviation of $\Delta I$ increase monotonically with $I$
(as expected) and the extended exponentials produce an excellent fit
for all bins. Specifically, for small $I$, the exponent of the fitted
exponential distribution is $n \approx 1$. This is because in this
case both $I^+$ and $I^-$ are close to zero and $\Delta I$ is
dominated by the estimation error, which falls off
exponentially~\cite{kw:Mar2005}. For large $I$, the estimation error
becomes smaller than the true mutual information difference between
the two random sub-samples,hence $n \approx 2$
from the central limit theorem.

\subsection{Interaction-specific modulator discovery}

Given a TF, $g_{\textit{\tiny{TF}}}$, and a set of candidate
modulators ${g_m}$ selected as previously discussed, we compute
$I(g_{\textit{\tiny{TF}}}, g_t)$ and $\Delta I(g_{\textit{\tiny{TF}}},
g_t | g_m)$ for all genes $g_t $ in the expression profile such that
$g_t \ne g_m$ and $g_t \ne g_{\textit{\tiny{TF}}}$. Significance of
each $\Delta I$ is then evaluated as a function of $I$, using the
extended exponentials from our null hypothesis model. Gene pairs with
a statistically significant p-value ($p < 0.05$), after Bonferroni
correction for multiple hypothesis testing, are retained for further
analysis.

\begin{figure}[htb]
    \centering
    \subfloat[ ]{
        \label{fig:fig3:a}
        \includegraphics[width=0.4\linewidth]{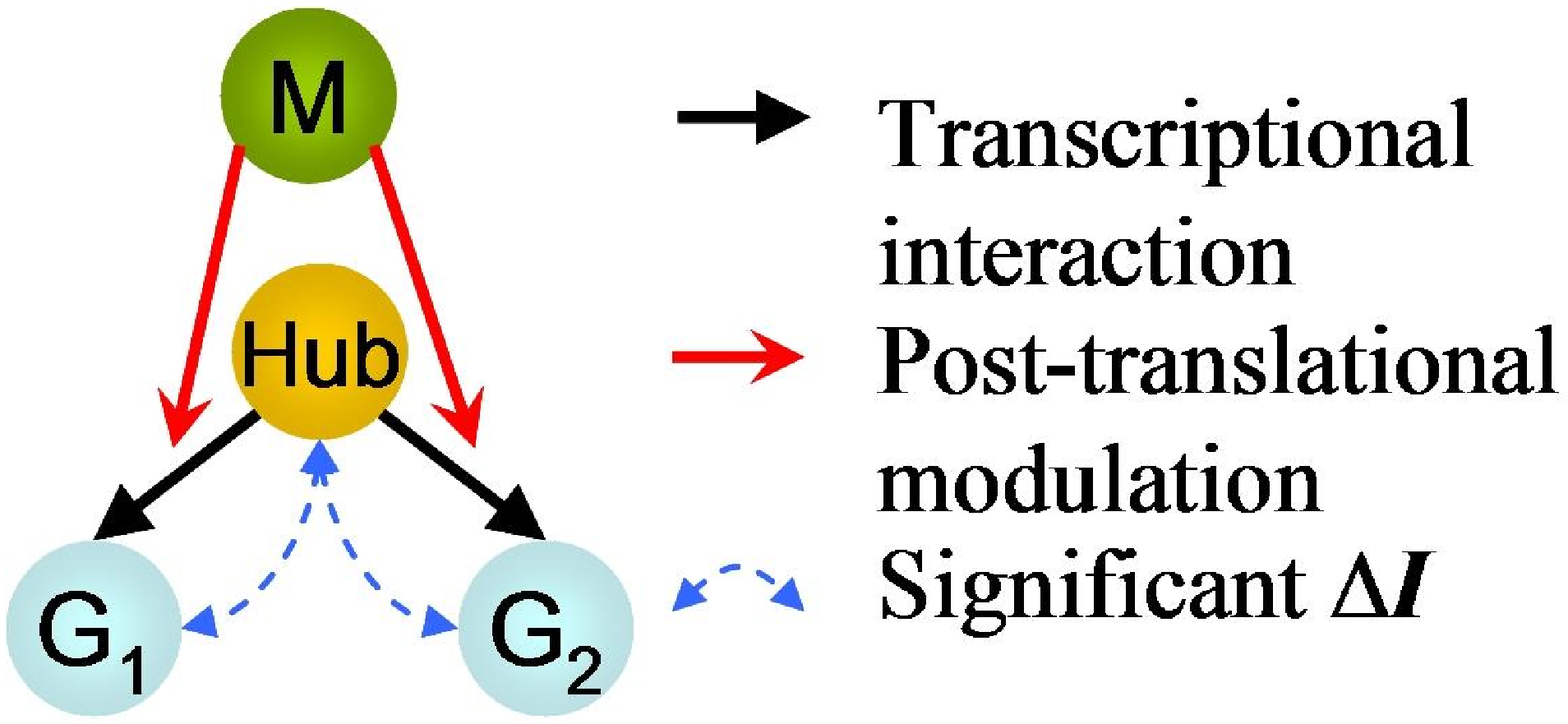}}
        \hspace{0.1\linewidth}
    \subfloat[ ]{
        \label{fig:fig3:b}
        \includegraphics[width=0.4\linewidth]{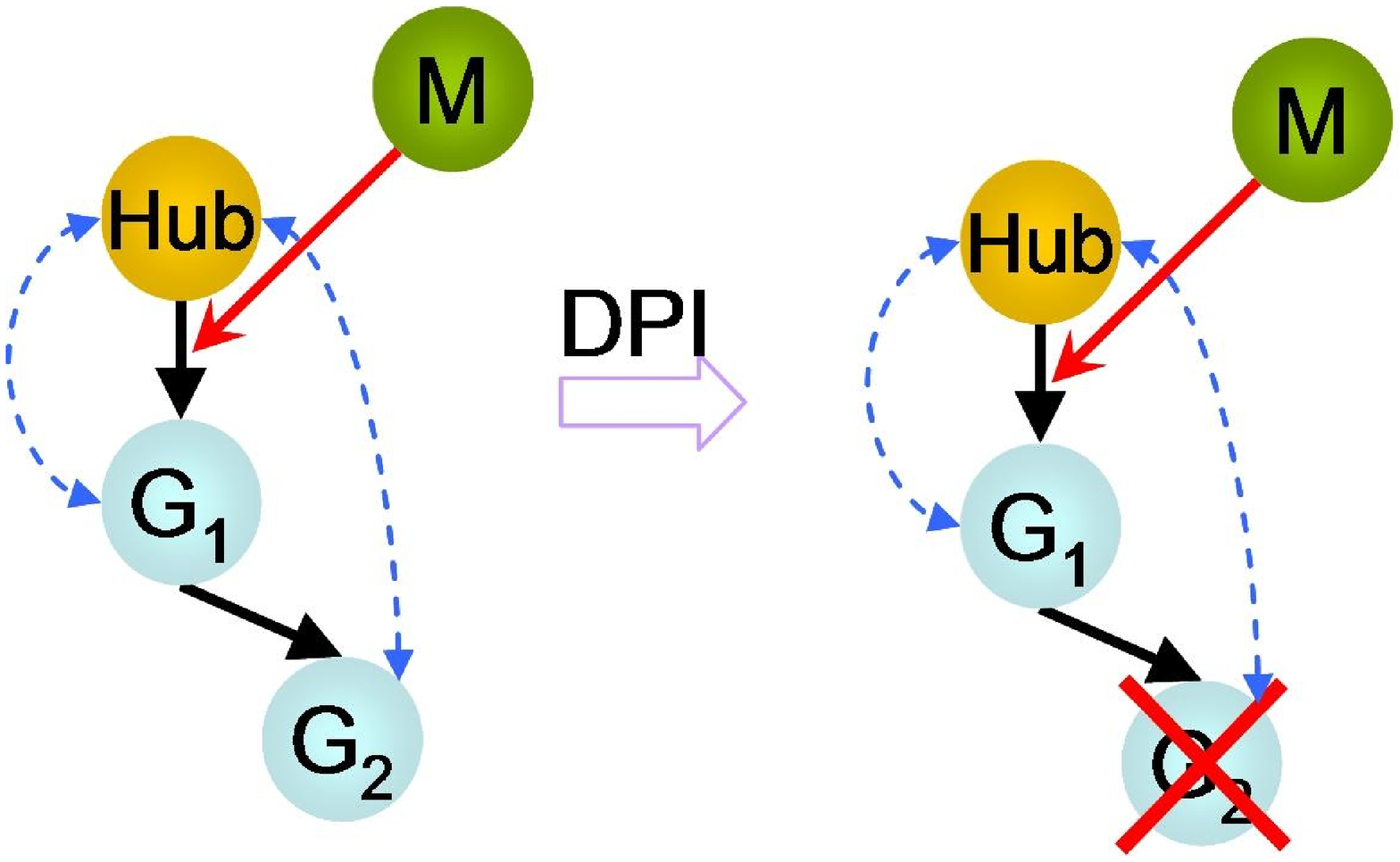}} \\
    \subfloat[ ]{
        \label{fig:fig3:c}
        \includegraphics[width=0.4\linewidth]{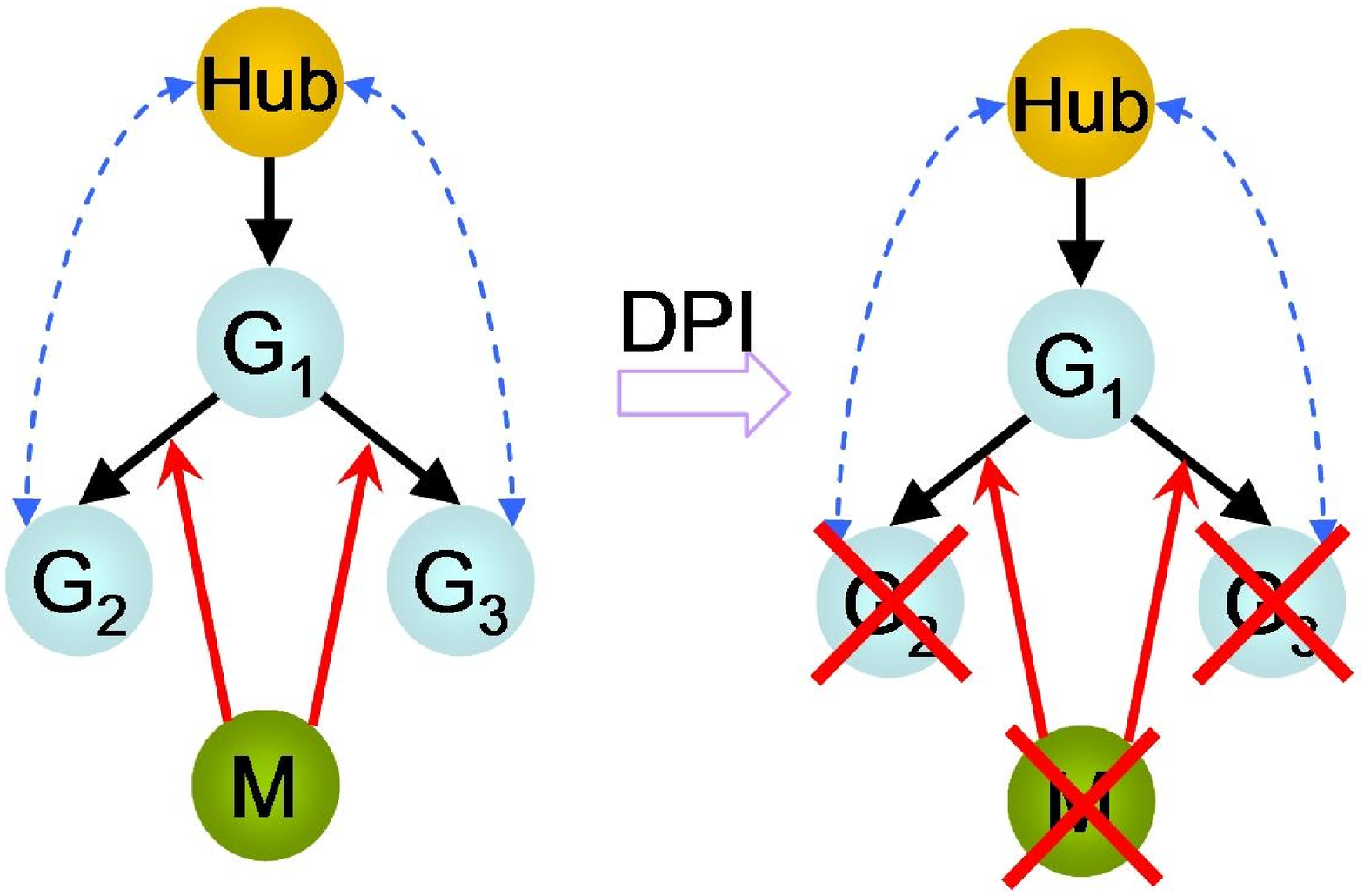}}
    \caption{Schematic diagram of the effect of DPI on eliminating
    indirect regulatory relationships.
    (a) Correct modulation model where the modulator (M) significantly
    changes the regulatory relationship between the TF (Hub) and its
    direct targets ($G_1$ and $G_2$).
    (b) Removal of indirect connections to the hub eliminates the
    detection of a significant $\Delta I$ on indirect targets.
    (c) Modulators that affect the downstream target of the TF hub,
    thus causing significant $\Delta I$ between the TF and its indirect
    neighbors, will be removed by applying DPI.}
    \label{fig:fig3}
\end{figure}

Significant pairs are further pruned if the interaction between
$g_{\textit{\tiny{TF}}}$ and $g_t$ is inferred as an indirect one in
both conditions $g_m^{\pm}$, based on the
ARACNE~\cite{kw:Bas2005,kw:Mar2005} analysis on the two subsets
$L_m^{\pm}$. This is accomplished by using the Data Processing
Inequality (DPI), a well-know property of MI introduced
in~\cite{kw:Bas2005,kw:Mar2005}, which states that the interaction
between $g_{\textit{\tiny{TF}}}$ and $g_t$ is likely indirect (i.e.
mediated through a third gene $g_x$), if $I(g_{\textit{\tiny{TF}}},
g_t) < min[ I(g_{\textit{\tiny{TF}}}, g_x), I(g_t, g_x) ]$. This step
eliminates some specific cases, illustrated in Fig.~\ref{fig:fig3},
where $g_m$ can produce a significant $\Delta I$ even though it does
not directly affect the $g_{\textit{\tiny{TF}}} \leftrightarrow g_t$
interaction. Briefly, two cases will be addressed by the use of the
DPI: (a) $g_m$ affects the $g_{\textit{\tiny{TF}}} \leftrightarrow
g_x$ interaction instead of $g_{\textit{\tiny{TF}}} \leftrightarrow
g_t$ (Fig.~\ref{fig:fig3:b}); and (b) $g_m$ modulates $g_x$, therefore
affecting the $g_x \leftrightarrow g_t$ interaction instead of the
$g_{\textit{\tiny{TF}}} \leftrightarrow g_t$. Thus $g_m$ is not a
modulator of the $g_{\textit{\tiny{TF}}}$ gene and should be removed
(Fig.~\ref{fig:fig3:c}). As discussed in~\cite{kw:Bas2005,kw:Mar2005},
the DPI was applied with a 15\% tolerance to minimize the impact of
potential MI estimation errors.

\section{RESULTS}
\subsection{Synthetic Model}
We first tested our approach on a simple synthetic network
(Fig.~\ref{fig:fig1:b}) that explicitly models two post-translational
modifications (activation by phosphorylation and by co-factor binding)
that modulate the ability of a TF to affect its targets. The synthetic
network includes a TF, a protein kinase (Ki) that phosphorylates the
TF, a co-factor (CF) that can bind to TF forming a transcriptionally
active complex, and three downstream targets of the TF's isoforms. The
transcription activation/inhibition was modeled using Hill kinetics
with exponential decay of mRNA molecules. Phosphorylation and cofactor
binding were modeled using Michaelis Menten and mass-action kinetics
respectively (see Supplementary Table 1 for kinetic equations).

A set of 250 synthetic expression profiles was generated from this
model using Gepasi (ver 3.30)~\cite{kw:Men97} by (a) randomly sampling
the independent variables (concentration of mRNA for the TF, Ki, and
CF) from a uniform distribution, so that they were statistical
independent (b) simulating network dynamics until a steady state was
reached, and (c) measuring the concentration of all mRNA species that
were explicitly represented in the network (using a Gaussian
experimental noise model with mean 0 and standard deviation equal to
10\% of the mean concentration for each variable). Note that only the
mRNA concentrations were used as inputs to our algorithm, even though
all molecular species (including all proteins isoforms) were
explicitly represented in the model. By conditioning on the expression
of the Ki and CF, using the 40\% of expression profiles with their
most and least expressed values, our approach correctly identified the
two and only two significant $\Delta I$, associated with the pairs
(CF, $g_2$) and (Ki, $g_3$), as shown in Fig.~\ref{fig:fig1:c}.

\subsection{Analysis of Human B lymphocyte data}
We then used our method to identify a genome-wide repertoire of
post-transcrip\-tio\-nal modulators of the MYC proto-oncogene -- a TF
which represents a major hub in the B cell transcriptional
network~\cite{kw:Bas2005}. The analysis was performed on a collection
of 254 gene expression profiles, representing 27 distinct cellular
phenotypes derived from populations of normal and neoplastic human B
lymphocytes. The gene expression profiles were collected using the
Affymetrix HG-U95A GeneChip\textregistered\ System (approximately
12,600 probes). Probes with absolute expression mean $\mu < 50$ and
standard deviation $\sigma < 0.3\mu$, were considered non-informative
and were excluded a-priori from the analysis, leaving 7907 genes.

We further selected 1117 candidate modulators with sufficient
expression range ($\mu > 200$ and $\sigma > 0.5\mu$) that were
statistically independent of MYC based on MI (significance was
established as in~\cite{kw:Mar2005}). The top 40\% and bottom 40\% of
the expression profiles in which a candidate modulator $g_m$ is
expressed at its highest and lowest levels, respectively, were used to
define the two conditional subsets $L_m^{\pm}$. The choice of the 40\%
threshold was specific to this dataset. It ensured that $\geq 100$
samples were available within each conditional subset for estimating
MI with a reasonable accuracy~\cite{kw:Mar2005}, while keeping the
modulators' expression range within the two subsets as separated as
possible.

The analysis inferred a repertoire of 100 genes, at a 5\% statistical
significance level (Bonferroni corrected), which are responsible for
modulating 205 regulatory relationships between MYC and its 130
inferred targets in an interaction-specific fashion. See Supplementary
Fig. 1 for a map of the modulators and the affected interactions. A
complete list is available in the Supplementary Table 2.

\subsection{Gene Ontology enrichment analysis}
To analyze the biological significance of these putative modulator
genes, we studied the enrichment of the Gene
Ontology~\cite{kw:Ash2000} Molecular Function categories among the 100
modulators compared to the initial list of 1117 candidate
modulators. As shown in Table~\ref{table:table1}, the top enriched
categories represent functions consistent with their activities as
modulators of cellular interactions. In particular, putative
modulators were enriched in kinases (PKN2, MAP4K4, BRD2, CSNK1D, HCK,
LCK, TRIB2, BRD2 and MARCKS), acyltransferase (GGT1, SAT and TGM2) and
transcriptional regulators (CUTL1, SSBP2, MEF2B, ID3, AF4, BHLHB2,
CREM, E2F5, MAX, NR4A1, CBFA2T3, REL, FOS and NFKB2). This is in
agreement with the established evidence that MYC is modulated through
phosphorylation and acetylation events, affecting its protein
stability~\cite{kw:Sea2000,kw:Pat2004}, and that MYC requires broadly
distributed effector proteins to influence its genomic
targets~\cite{kw:Lev2003}. We also found that 4 of the 6 modulators
with the largest number of affected targets (e.g. UBE2G1, HCK, USP6
and IFNGR1), are associated with non-target-specific functions
(e.g. protein degradation, upstream signaling pathway components and
receptor signaling molecules, etc). On the other hand, the 14
modulators that are transcription factors (and may thus be MYC
co-factors) tend to be highly interaction-specific, affecting only 1-4
target genes (see Supplementary Fig. 1).

\begin{table}
    \centering
    \caption{Most enriched Gene Ontology Molecular Function categories
    for the inferred MYC modulators. False discovery rate (FDR) are calculated
    from Fisher's exact test and adjusted for multiple hypothesis testing.
    Only categories with at least 5 genes from the initial 1117 candidate
    modulators were used.}
    \begin{tabular}{l@{\hspace{1in}}c}
    \hline
    Gene Ontology Molecular Function Categories & Enrichment FDR \\
    \hline
    DNA binding                                 & 0.007 \\
    Transferase activity                        & 0.010 \\
    Acyltransferase activity                    & 0.010 \\
    Antioxidant activity                        & 0.018 \\
    Phosphoric monoester hydrolase activity     & 0.026 \\
    Adenyl nucleotide binding                   & 0.028 \\
    Transcription regulator activity            & 0.052 \\
    Protein serine/threonine kinase activity    & 0.066 \\
    \hline
    \end{tabular}
    \label{table:table1}
\end{table}

\subsection{Literature validation of known MYC modulators}
Closer scrutiny, through literature review reveals that a number of
the inferred modulators play a role in the post-transcriptional and
post-translational modulation of MYC, either by direct physical
interaction, or by modulating well-characterized pathways that are
known to affect MYC function.

Among the list of putative modulators, we found two well known
co-factors of MYC: MAX and MIZ-1. Numerous studies,~\cite{kw:Ama93}
among many others, have shown that transcriptional activation by MYC
occurs via dimerization with its partner MAX. Similarly, MIZ-1 has
been shown to specifically interact with MYC through binding to its
helix-loop-helix domain, which may be involved in gene repression by
MYC~\cite{kw:Peu97}. Several protein kinases identified by our method
are also notable: CSNK1D, a member of the Casein Kinase I gene family,
is a reasonable MYC modulator since one of its related family member,
Casein Kinase II, has been demonstrated to phosphorylate MYC and MAX,
thus affecting the DNA-binding kinetics of their
heterodimer~\cite{kw:Lus89,kw:Bou93}. MYC is also know to be
phosphorylated by JNK~\cite{kw:Nog99} and GSK~\cite{kw:Gre2003}, which
affect the stability of its protein. Although both kinases were
excluded from our initial candidate modulator set due to their
insufficient expression range, our approach was able to identify some
of their upstream signaling molecules, such as MAP4K4 and HCK.  Both
MAP4K4 and HCK are members of the BCR singling pathway that is known
to control MYC activation and degradation~\cite{kw:Nii2002}.  In
particular, MAP4K4 has been previously reported to specifically
activate JNK~\cite{kw:Mac2004}.

MYC stability is also known to be regulated through ubiquitin-mediated
proteolysis~\cite{kw:Sal99}. Two enzymes in this process, USP6 and
UBE2G1, were identified as putative modulators of MYC. Although there
is no biochemical evidence implicating these two proteins
specifically, they serve as a reasonable starting point for
biochemical validation.  We also identified putative modulators that
could potentially influence the MYC mRNA stability. One of them,
APOBEC3B, is closely related to APOBEC1, which has been well
characterized as a RNA-editing enzyme capable of binding MYC mRNA in
the 3' untranslated region, thus increasing its
stability~\cite{kw:Ana2000}. While APOBEC1 was excluded from our
analysis due to its insufficient expression range, the identification
of its closely related family member, APOBEC3B, may suggest a similar
mechanism. Another protein from this category, HNRPDL, encodes a
heterogeneous nuclear ribonucleoprotein which interacts with mRNA and
may have a role in mRNA nuclear export.

MYC stimulates gene expression in part at the level of chromatin,
through its association with co-factors that affect the histone
acetylation and DNA methylation. DNMT1, which encodes a DNA
methyltransferase, was found in our putative modulator list. Current
literature suggests that MYC may repress transcription through the
recruitment of DNA methyltransferase as corepressor, which may in turn
lead to hypoacetylated histones that are often associated with
transcriptional silencing~\cite{kw:Bre2005,kw:Rob2000}.

Many other genes in our list of putative modulators of MYC also
present relevant biological functionality, such as transcription
factors FOS, CREM, REL and NFKB2, anti-apoptosis regulator BCL-2, to
name but a few. Those for which functional relevance can not be
established from the current literature likely belong to two groups:
(a) novel bona fide MYC modulators requiring further biochemical
validation and (b) genes that are co-expressed with a bona fide
modulator, such as gene from the same biological pathway. A likely
example of the latter case is NFKB2 and its inhibitor NFKBIA, which
are both identified as modulators of MYC, while having substantially
correlated expression profiles (Pearson correlation 0.55).

\subsection{Transcription factors co-binding analysis}
For the putative modulators annotated as TF, one potential mechanism
of modulation is as MYC co-factors. We thus searched for the binding
signatures of both MYC and the modulator-TF within the promoter region
of the genes whose interaction with MYC appeared to be modulated by
the TF. Of the 14 TFs in our putative modulator list, 8 have credibly
identified DNA binding signatures from TRANSFAC~\cite{kw:Win2001}
(represented as position-specific scoring matrix). These TFs affect 12
MYC interactions with 11 target genes.  Additionally, 4 of the 5
target genes whose expressions are positively correlated with MYC
present at least one E-Box in their promoter region ($p <
4.03\times10^{-4}$)\footnote{MYC is known to transcriptional activate
  its targets through binding to E-box elements. Repression by MYC
  occurs via a distinct mechanism, not involving E-boxes, which is not
  yet well characterized.} . As is shown in Table~\ref{table:table2},
of the 12 instances of statistically significant modulator target
pairs, 7 target genes harbor at least one high specificity TF binding
signature ($P_{BS} < 0.05$) in their promoter region. The overall
p-value associated with this set of events is $p < 0.0025$ (from the
binomial background model). This strongly supports the hypothesis that
these TFs are target specific co-factors of MYC.

\begin{table}[htb]
    \centering
    \caption{Promoter analysis of the MYC target genes affected by
      TF-modulators. Binding signatures of 8 of the 14 TFs in the
      putative modulator list were obtained through TRANSFAC.
      Promoter sequences of the target genes (2Kb upstream and
      2Kb downstream of the transcription initiation site)
      were retrieved from the UCSC Golden Path database~\cite{kw:Kar2003} and
      masked for repetitive elements. Statistical significance is
      assessed by considering a null score distribution computed
      from random sequences using an order-2 Markov model learned
      from the actual promoter sequences, where $P_{BS}$ is calculated
      as the probability of finding at least one binding site per
      1Kb sequence under the null hypothesis. We used a significance
      threshold of 0.05; findings below this threshold are shown in red.}
    \begin{tabular}{|c|c|c|c|}
    \hline
    Modulator & Binding Signature & MYC targets & $P_{BS}$ \\
    \hline
    CUTL1   & \includegraphics[width=0.2\linewidth]{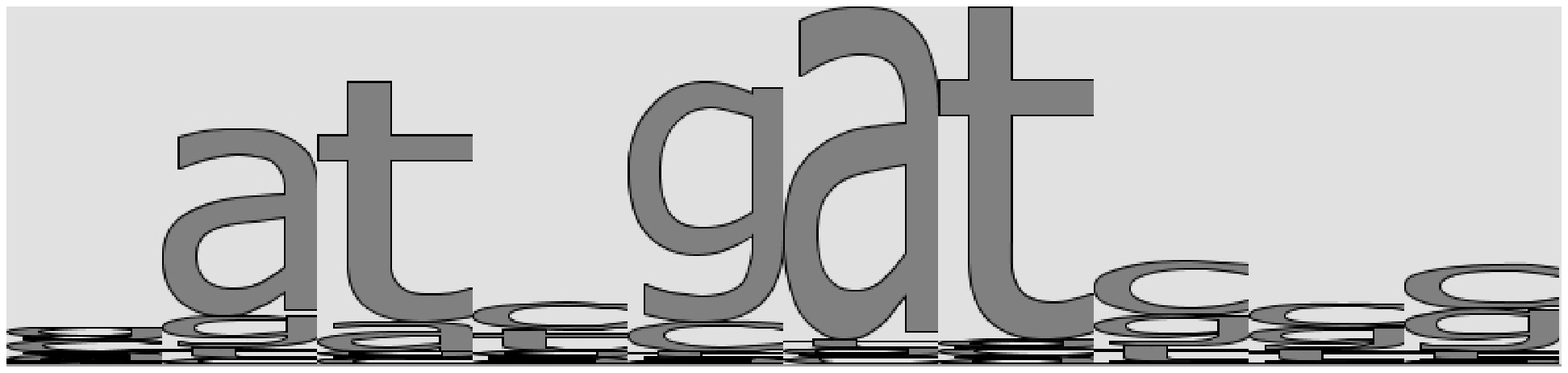}  & NP    & \textcolor{red}{0.032} \\
    \hline
    CREM    & \includegraphics[width=0.2\linewidth]{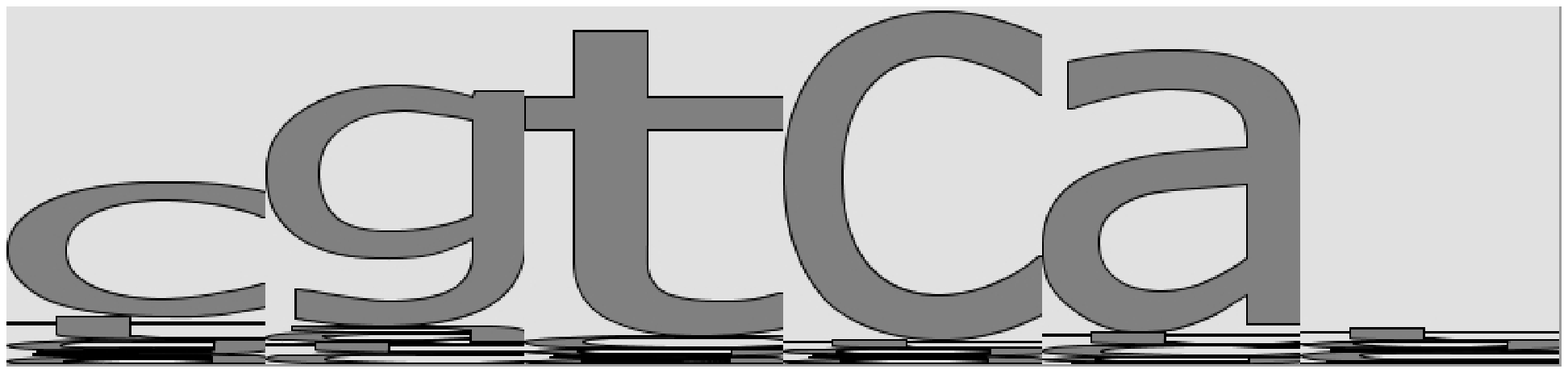}   & PRKDC & \textcolor{red}{0.041} \\
    \hline
    \multirow{3}{*}{BHLHB2} &
    \multirow{3}{*}{\includegraphics[width=0.2\linewidth]{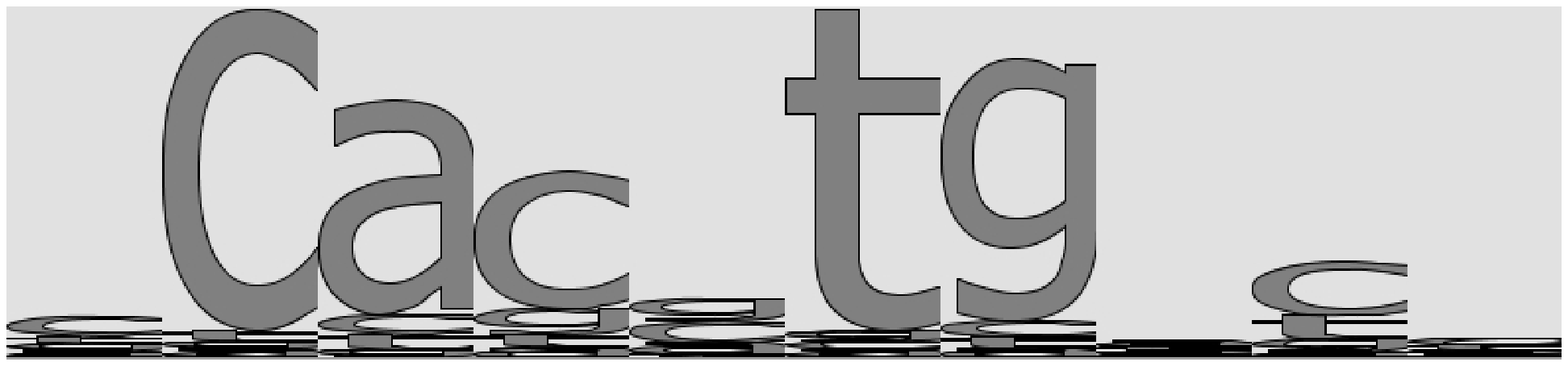}}  & TLE4  & \textcolor{red}{0.030} \\
    & & MLL   & \textcolor{red}{0.039} \\
    & & IL4R  & 0.140 \\
    \hline
    \multirow{3}{*}{MEF2B} &
    \multirow{3}{*}{\includegraphics[width=0.2\linewidth]{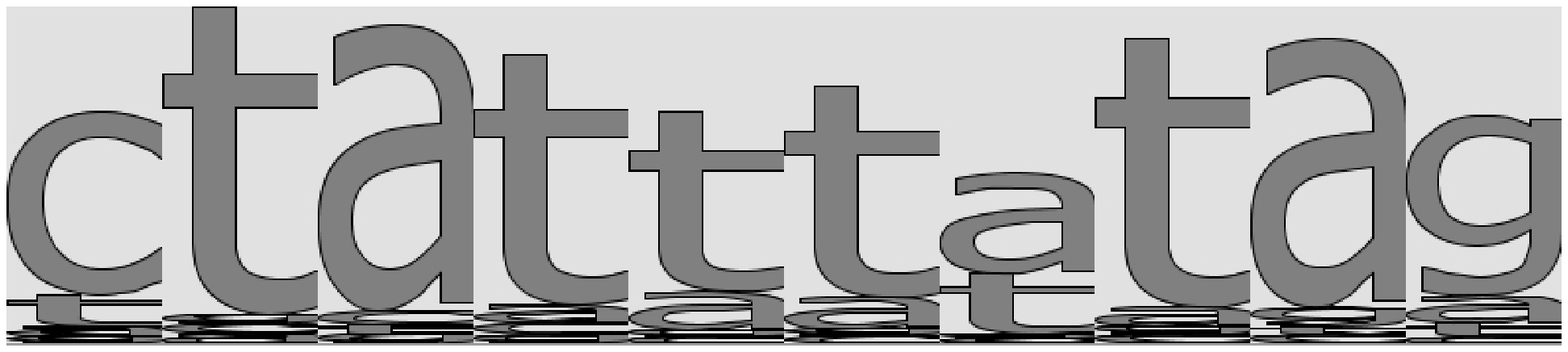}}   & KLF12  & 0.391 \\
    & & CR2   & 0.326 \\
    & & CYBB  & \textcolor{red}{0.045}\\
    \hline
    FOS     & \includegraphics[width=0.2\linewidth]{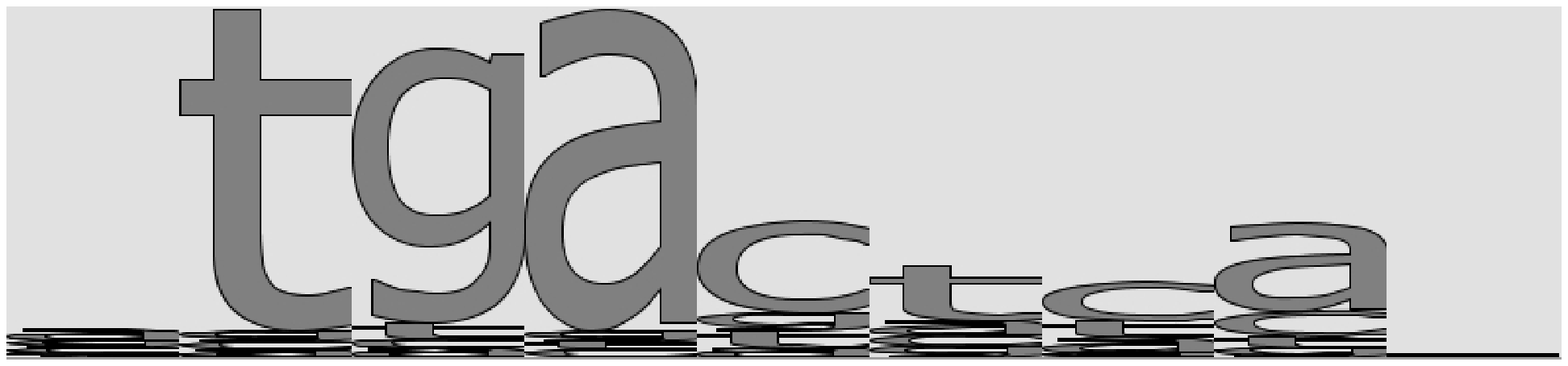}    & ZNF259    & \textcolor{red}{0.049} \\
    \hline
    REL     & \includegraphics[width=0.2\linewidth]{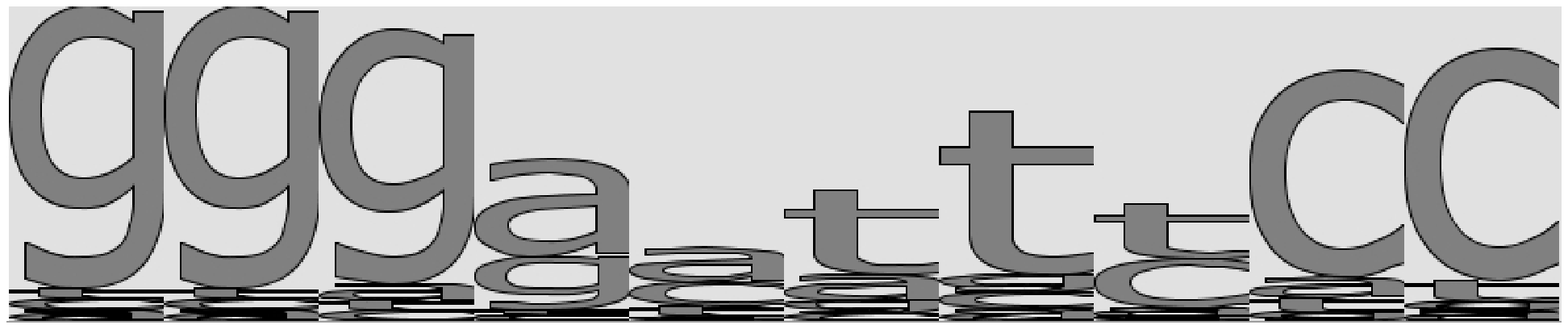}    & KEL       & 0.125 \\
    \hline
    E2F5    & \includegraphics[width=0.2\linewidth]{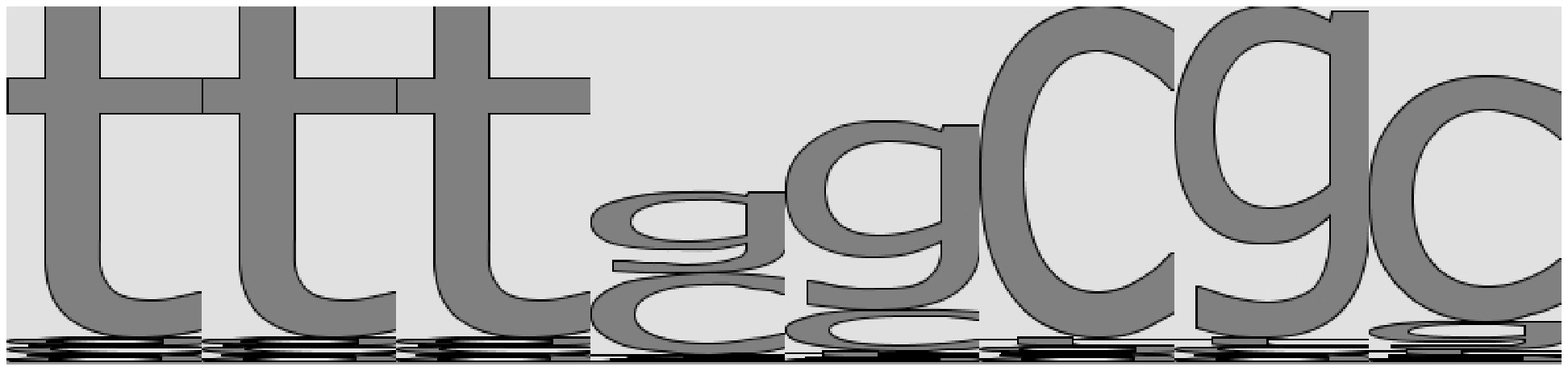}   & IL4R      & 0.870 \\
    \hline
    NFKB2   & \includegraphics[width=0.2\linewidth]{REL}    & FOXK2     & \textcolor{red}{0.041} \\
    \hline
    \end{tabular}
    \label{table:table2}
\end{table}

\section{CONCLUSION AND DISCUSSION}
Cellular interactions can be neither represented as a static
relationship nor modeled as pure pairwise processes. The two issues
are deeply interlinked as higher order interactions are responsible
for the rewiring of the cellular network in a context dependent
fashion. For transcriptional interactions, one can imagine a
transistor-like model, in which the ability of a TF to activate or
repress the expression of its target genes is modulated, possibly in a
target-specific way, by one or more signaling proteins or
co-factors. Such post-transcriptional and post-translational
conditional interactions are necessary to create complex rather than
purely reactive cellular behavior and should be abundant in
biological systems.

Unfortunately, most post-translational interactions (e.g.,
phosphorylation or complex formation) do not affect the mRNA
concentration of the associated proteins. As a result, they are
invisible to na\"{\i}ve co-expression analysis methods. However,
proteins that are involved in post-translational regulation may be
themselves transcriptionally regulated. At steady state, the
concentration of such post-translationally modified proteins and
complexes can then be expressed as a function of some mRNA
expressions, albeit in a non-obvious, conditional fashion. With this
in mind, we show that conditional analysis of pairwise statistical
dependencies between mRNAs can effectively reveal a variety of
transient interactions, as well as their post-transcriptional and
post-translational regulations.

In this paper, we restrict our search to genes that affect the ability
of a given TF to transcriptionally activate or repress its
target(s). While the identification of the targets of a transcription
factor is a rather transited area, the identification of upstream
modulators is essentially unaddressed at the computational level,
especially in a TF-target interaction specific way. Experimentally, it
constitutes an extremely complex endeavor that is not yet amenable to
high-throughput approaches. For instance, while hundreds of MYC
targets are known, only a handful of genes have been identified that
are capable of modifying MYC's ability to activate or repress its
targets. Even fewer of these are target specific.

We show that such modulator genes can be accurately and efficiently
identified from a set of candidates using a conditional MI difference
metric. One novelty of our approach is that it requires no a-priori
selection of the modulator genes based on certain functional criteria:
the candidate modulators include all genes on the microarray that have
sufficient dynamic range and no significant MI with the TF gene. The
first requirement can be actually lifted without affecting the method
other than making it more computationally intensive and requiring more
stringent statistical tests (as the number of tested hypotheses would
obviously increases). This is because conditioning on a gene with an
expression range comparable to the noise is equivalent to random
sub-sampling of the expression profiles, an event that will be
filtered out by our statistical test.

Another critical element of our method is the phenotypic heterogeneity
of the expression profiles. This ensures that no sufficiently large
subset of the expression profiles can be obtained without sampling
from a large number of distinct phenotypes, including both normal and
malignant cells. In fact, the average number of distinct cellular
phenotypes in any subset of gene expression profiles used in the
analysis is about 20, with no subset containing fewer than 13. Thus,
the modulators identified in this paper are not associated with a
specific cellular phenotype.

To derive a null model for estimating the significance of individual
conditional mutual information differences, $\Delta I$, we
investigated the statistics of $\Delta I$ as a function of the
unconditional mutual information $I$. Other models, such as dependence
of $\Delta I$ on $I^+$ or $I^-$, were also investigated, but they
proved to be less informative. It is possible that a more accurate
null model may be learned by studying the variation of $\Delta I$ as a
function of both $I$ and either $I^+$ or $I^-$. For example, this may
answer questions such as: given measured values of $I$ and $I^-$, what
is the probability of seeing a specific difference in $\Delta I$?
While this may provide finer-grained estimates of the statistical
significance, this also would dramatically increases the number of
Monte Carlo samples necessary for achieving a reasonable numerical
precision, which would prohibit the actual deployment of the strategy.

Method limitations follow broadly into three categories: (a) The
computational deconvolution of molecular interaction is still
manifestly inaccurate. This has obvious effects on the discovery of
interaction-specific modulators, i.e. we may identify modulators of
``functional'' rather than physical interactions. (b) The method
cannot dissect modulators that are constitutively expressed
(housekeeping genes) and activated only at the post-translational
level (e.g., the p53 tumor suppressor gene), nor modulators that are
expressed at very low concentrations. However, in both cases a gene
upstream of the most direct modulator may be identified in its
place. For instance, JNK is a known modulator of MYC activity, which
is weakly expressed in human B cells and, therefore it is not even
included in the initial candidate modulator list. However, MAP4K4,
which is upstream of JNK in the signaling cascade, is identified as a
MYC modulator in its place. (c) The method cannot disambiguate true
modulators from those co-expressed with them.

Techniques to deal with all these drawbacks are currently being
investigated. However, we believe that, even in its current state, our
approach presents a substantial advancement in the field of reverse
engineering of complex cellular networks.

\section{SUPPLEMENTARY MATERIAL}
Supplementary Materials are available at:
\url{http://www.dbmi.columbia.edu/~kaw7002/recomb06/supplement.html}.

\section{ACKNOWLEDGEMENT}

We thank R. Dalla-Favera, K. Basso and U. Klein for sharing the B
Cell gene expression profile dataset and helpful discussions.

% ---- Bibliography ----
%

%

\begin{thebibliography}{99}
\bibitem {kw:Fri2004} Friedman, N. Inferring cellular networks using
  probabilistic graphical models. {\em Science} {\bf 303} (2004)
  799--805.

\bibitem {kw:Gar2003} Gardner, T.~S.\ and di Bernardo, D.\ and Lorenz,
  D., Collins, J.~J.: Inferring genetic networks and identifying
  compound mode of action via expression profiling. {\em Science} {\bf
    301} (2003) 102--105.

\bibitem {kw:Elk2003} Elkon, R., Linhart, C., Sharan R., Shamir, R.,
  Shiloh, Y.: Genome-Wide In Silico Identification of Transcriptional
  Regulators Controlling the Cell Cycle in Human Cells. {\em Genome
    Res.} {\bf 13} (2003) 773--780.

\bibitem {kw:Stu2003} Stuart, J.~M., Segal, E., Koller, D., Kim,
  S.~K.: A gene-coexpression network for global discovery of conserved
  genetic modules. {\em Science} {\bf 302} (2003) 249--55.

\bibitem {kw:Bas2005} Basso, K., Margolin, A.~A., Stolovitzky, G.,
  Klein, U., Dalla-Favera, R., Califano, A: Reverse engineering of
  regulatory networks in human B cells. {\em Nat.\ Gen.} {\bf 37}
  (2005) 382--390.

\bibitem {kw:Zei2003} Zeitlinger, J., Simon, I., Harbison, C.~T.,
  Hannett, N.~M., Volkert, T.~L., Fink, G.~R., Young, R.~A.:
  Program-Specific Distribution of a Transcription Factor Dependent on
  Partner Transcription Factor and MAPK Signaling. {\em Cell} {\bf
    113} (2003) 395--404.

\bibitem {kw:Lus2004} Luscombe, N.~M., Babu, M.~M., Yu, H., Snyder,
  M., Teichmann, S.~A., Gerstein, M.: Genomic analysis of regulatory
  network dynamics reveals large topological changes. {\em Nature}
  {\bf 431} (2004) 308--12.

\bibitem {kw:Seg2003} Segal, E., Shapira, M., Regev, A., Pe'er, D.,
  Botstein, D., Koller, D., Friedman, N.: Module networks: identifying
  regulatory modules and their condition-specific regulators from
  expression data. {\em Nat.\ Gen.} {\bf 34} (2003) 166--176.

\bibitem {kw:deL2005} de Lichtenberg, U., Jensen, L.~J., Brunak, S.,
  Bork, P.: Dynamic Complex Formation During the Yeast Cell
  Cycle. {\em Science} {\bf 307} (2005) 724--727.

\bibitem {kw:Pe'2002} Pe'er, D., Regev, A., Tanay, A.: Minreg:
  Inferring an active regulator set. {\em Bioinformatics} {\bf 18}
  (2002) S258--S267.

\bibitem {kw:Mar2005} Margolin, A., Nemenman, I., Basso, K., Klein,
  U., Wiggins, C., Stolovitzky, G., Dalla-Favera, R., Califano, A.:
  ARACNE: An algorithm for reconstruction of genetic networks in a
  mammalian cellular context. {\em BMC Bioinformatics} (2006). In
  press.  (manuscript available online at
  \url{http://arxiv.org/abs/q-bio.MN/0410037)}.

\bibitem {kw:Nem2004} Nemenman, I.: Information theory, multivariate
  dependence, and genetic network inference KITP, UCSB,
  NSF-KITP-04-54, Santa Barbara, CA (2004) (manuscript available
  online at \url{http://arxiv.org/abs/q-bio/0406015}).

\bibitem {kw:But2000} Butte, A.~J., Kohane, I.~S.: Mutual information
  relevance networks: functional genomic clustering using pairwise
  entropy measurements.  {\em Pac.\ Symp.\ Biocomput.} (2000) 418--29.

\bibitem {kw:Fri2000} Friedman, N., Linial, M., Nachman, I., Pe'er,
  D.: Using Bayesian networks to analyze expression data. {\em J.\
    Comp.\ Biol.} {\bf 7} (2000) 601--620.

\bibitem {kw:Men97} Mendes, P.: Biochemistry by numbers: simulation of
  biochemical pathways with Gepasi 3. {\em Trends Biochem.\ Sci.} {\bf
    22} (1997) 361--363.

\bibitem {kw:Ash2000} Ashburner, M. et al.: Gene Ontology: tool for
  the unification of biology. {\em Nat. Gen.} {\bf 25} (2000)
  1061--4036.

\bibitem {kw:Sea2000} Sears, R., Nuckolls, F., Haura, E., Taya, Y.,
  Tamai, K., Nevins, J.~R.: Multiple Ras-dependent phosphorylation
  pathways regulate Myc protein stability. {\em Genes Dev.} {\bf 14}
  (2000) 2501--2514.

\bibitem {kw:Pat2004} Patel, J.~H. et al.: The c-MYC Oncoprotein Is a
  Substrate of the Acetyltransferases hGCN5/PCAF and TIP60. {\em Mol.\
    Cell.\ Biol.} {\bf 24} (2004) 10826--10834.

\bibitem {kw:Lev2003} Levens, D. L.: Reconstructing MYC. {\em Genes
    Dev.}  {\bf 17} (2003) 1071--1077.

\bibitem {kw:Ama93} Amati, B., Brooks, M. W., Levy, N., Littlewood,
  T.~D., Evan, G.~I., Land, H: Oncogenic activity of the c-Myc protein
  requires dimerization with Max. {\em Cell} {\bf 72} (1993) 233--245.

\bibitem {kw:Peu97} Peukert, K. et al.: An alternative pathway for
  gene regulation by Myc. {\em EMBO J.} {\bf 16} (1977) 5672--5686.

\bibitem {kw:Lus89} Luscher, B., Kuenzel, E.~A., Krebs, E.~G.,
  Eisenman, R.~N.: Myc oncoproteins are phosphorylated by casein
  kinase II. {\em EMBO J.} {\bf 8} (1989) 1111--1119.

\bibitem {kw:Bou93} Bousset, K., Henriksson, M., Luscher-Firzlaff,
  J.~M., Litchfield, D.~W., Luscher, B: Identification of casein
  kinase II phosphorylation sites in Max: effects on DNA-binding
  kinetics of Max homo- and Myc/Max heterodimers. {\em Oncogene} {\bf
    8} (1993) 3211--3220.

\bibitem {kw:Nog99} Noguchi, K. et al.: Regulation of c-Myc through
  Phosphorylation at Ser-62 and Ser-71 by c-Jun N-Terminal
  Kinase. {\em J.\ Biol.\ Chem.} {\bf 274} (1999) 32580--32587.

\bibitem {kw:Gre2003} Gregory, M.~A., Qi, Y., Hann, S.~R.:
  Phosphorylation by glycogen synthase kinase-3 controls c-myc
  proteolysis and subnuclear localization. {\em J.\ Biol.\ Chem.} {\bf
    278} (2003) 51606--51612

\bibitem {kw:Nii2002} Niiro, H., Clark, E.~A.: Regulation of B-cell
  fate by antigen-receptor signals. {\em Nat.\ Rev.\ Immun.} {\bf 2}
  (2002) 945--956.

\bibitem {kw:Mac2004} Machida, N. et al.: Mitogen-activated Protein
  Kinase Kinase Kinase Kinase 4 as a Putative Effector of Rap2 to
  Activate the c-Jun N-terminal Kinase. {\em J.\ Biol.\ Chem.} {\bf
    279} (2004) 15711--15714.

\bibitem {kw:Sal99} Salghetti, S.~E., Kim, S.~Y., Tansey, W.~P.:
  Destruction of Myc by ubiquitin-mediated proteolysis:
  cancer-associated and transforming mutations stabilize Myc. {\em
    EMBO J.} {\bf 18} (1999) 717--726.

\bibitem {kw:Ana2000} Anant, S., Davidson, N.~O.: An AU-Rich Sequence
  Element (UUUN[A/U]U) Downstream of the Edited C in Apolipoprotein B
  mRNA Is a High-Affinity Binding Site for Apobec-1: Binding of
  Apobec-1 to This Motif in the 3' Untranslated Region of c-myc
  Increases mRNA Stability. {\em Mol.\ Cell.\ Biol.} {\bf 20} (2000)
  1982--1992.

\bibitem {kw:Bre2005} Brenner, C. et al.: Myc represses transcription
  through recruitment of DNA methyltransferase corepressor. {\em EMBO
    J.} {\bf 24} (2005) 336--346.

\bibitem {kw:Rob2000} Robertson, K.~D. et al. DNMT1 forms a complex
  with Rb, E2F1 and HDAC1 and represses transcription from
  E2F-responsive promoters.  {\em Nat.\ Gen.} {\bf 25} (2000)
  338--342.

\bibitem {kw:Win2001} Wingender, E. et al.: The TRANSFAC system on
  gene expression regulation. {\em Nucl.\ Acids Res.} {\bf 29} (2001)
  281--283.

\bibitem {kw:Kar2003} Karolchik, D. et al.: The UCSC Genome Browser
  Database. {\em Nucl.\ Acids Res.} {\bf 31} (2003) 51--54.

\end{thebibliography}
\end{document}